\documentclass[pre,showpacs,superscriptaddress,nofootinbib]{revtex4-1}

\usepackage{graphicx}% Include figure files
\usepackage{bm}% bold math
\usepackage{fleqn}
\usepackage{amssymb}
\usepackage{color}
\usepackage{epstopdf}
\usepackage{amsmath}
\usepackage{amstext}
\usepackage{latexsym}
\usepackage[colorlinks=true,citecolor=blue,linkcolor=blue]{hyperref}
\usepackage{mathptmx} %comapct fonts
\newcommand{\beq}{\begin{equation}}
\newcommand{\eeq}{\end{equation}}
\newcommand{\beqa}{\begin{eqnarray}}
\newcommand{\eeqa}{\end{eqnarray}}

\begin{document}

\title{Delocalization and heat transport in multidimensional trapped ion systems}

\author{A. Ruiz-Garc\'\i a}
\affiliation{Departamento de F\'isica, Universidad de La Laguna, La Laguna 38203, Spain}
\affiliation{IUdEA Instituto Universitario de Estudios Avanzados, Universidad de La Laguna, La Laguna 38203, Spain}
\author{J.J. Fern\'andez}
\affiliation{Astrophysics Research Institute, Liverpool John Moores University, 146 Brownlow Hill, Liverpool L3 5RF, UK}
\author{D. Alonso}
\affiliation{Departamento de F\'isica, Universidad de La Laguna, La Laguna 38203, Spain}
\affiliation{IUdEA Instituto Universitario de Estudios Avanzados, Universidad de La Laguna, La Laguna 38203, Spain}

\begin{abstract}
We study the connection between heat transport properties of systems coupled to different thermal baths in two separate regions and their underlying nonequilibrium dynamics. We consider classical systems of interacting particles that may exhibit a certain degree of delocalization, and whose effective dimensionality can be modified through the controlled variation of a global trapping potential. We focus on Coulomb crystals of trapped ions, which offer a versatile playground to shed light on the role that spatial constraints play on heat transport. We use a three-dimensional model to simulate the trapped ion system, and show in a numerically rigorous manner to what extent heat transport properties could be feasibly tuned across the structural phase transitions between the linear, planar zigzag and helical configurations. By solving the classical Langevin equations of motion, we analyze the steady state spatial distributions of the particles, the temperature profiles and total heat flux through the various structural phase transitions that the system may experience. The results evidence a clear correlation between the degree of delocalization of the internal ions and the emergence of a non-zero gradient in the temperature profiles. The signatures of the phase transitions in the total heat flux as well as the optimal spatial configuration for heat transport are shown.
\end{abstract}

\pacs{05.60.-k,64.60.Ht, 05.70.Fh, 37.10.Ty}
%05.60.-k:	Transport processes
%05.60.Cd:	Classical transport
%64.60.Ht: Dynamic critical phenomena
%05.70.Fh: Phase transitions: general studies
%37.10.Ty : ion trapping

\maketitle

\section{Introduction}

The downsizing of electronic devices to the nanometric scale, driven by the rapid progress of micro-electronic technology, has made the problem of thermal conduction increasingly important because of the need to find ways to dissipate a significant amount of energy in a shrinking compact space \cite{dubi11}. In this sense, the divergence of thermal conductivity with the size of the materials of reduced dimensionality would allow the construction of nano-materials capable of dissipating heat efficiently. This would solve one of the fundamental problems arising from the miniaturization of electronic and optical devices. 

According to Fourier's law, given a system connected to different heat reservoirs in two separate regions, the amount of heat transferred per unit area and time unit has a linear dependence on the imposed temperature gradient, the thermal conductivity that characterizes the material being the constant of proportionality between both magnitudes. Although this law can be verified in a simple way for three-dimensional systems, it is known that heat conduction exhibits anomalous behavior in systems of reduced dimensionality, such as carbon nanotubes, silicon nanowires or molecular junctions \cite{dubi11,sevik11,lee17}. This issue keeps many important and fundamental questions open in the field of nonequilibrium statistical physics \cite{lepri03,dhar08,gaspard08}.

From a theoretical perspective, the study of heat transport through nanoscale devices typically requires making tradeoffs between the size of the system and the completeness of the model. Most of the work on low-dimensional systems have considered simple, yet nontrivial, models that incorporate elements crucial to heat conduction, such as anharmonicity and disorder \cite{li01,dhar08b,lepri03,dhar08}. It has been shown that some one-dimensional systems, such as the Frenkel-Kontorova model \cite{hu98} or the Lorentz model \cite{alonso99}, the temperature gradient is uniform and the thermal conductivity is a constant, independent of the size of the system. This indicates that these systems obey Fourier's law. In contrast, in the case of one-dimensional integrable systems, such as a harmonic chain and the Toda mono-atomic model, a temperature gradient is not established \cite{rieder67,dhar16,toda79,kundu16}. There are also one-dimensional nonintegrable systems, such as the Fermi-Pasta-Ulam model \cite{kaburaki93,lepri97,lepri98} or the Toda diatomic chain \cite{hatano99}, for which the thermal conductivity diverges with increasing system size. In two dimensions, anomalous conductivities exhibiting a logarithmic divergence with the size of system have been reported \cite{lepri03,dhar08,yang06}. In the case of polygonal channels with zero Lyapunov exponents, numerical simulations have shown transport properties ranging from normal to anomalous conduction, depending on the system parameters \cite{li02,alonso02,alonso04}. Although these mathematical models have shed some light on the underlying mechanisms for normal heat conduction, understanding heat transport at the microscopic level remains a central topic of current research.

Coulomb crystals of ions confined in electromagnetic traps and manipulated with laser beams \cite{ghosh95} provide a versatile platform to study a broad range of intriguing physical phenomena emerging in system driven out of equilibrium, in particular energy transport in both classical and quantum regimes \cite{pruttivarasin11,lin11,bermudez13,ramm14,ruiz14,freitas15,cormick16,blatt12,bermudez16,cosco17}. Due to the unique control in the preparation, manipulation, and detection of the electronic and motional degrees of freedom, they have also become a promising candidate for quantum information and computation, quantum networks, quantum simulations, and quantum metrology \cite{haffner08,schneider12,pezze18}. From the theoretical point of view, Coulomb crystals are particularly interesting since the typical distances between ions of several micrometers, with densities several orders of magnitude smaller than the standard crystalline materials, simplifies to a great extent the analysis of many issues related to their very rich and highly non-trivial static and dynamical properties.

It has been shown that the thermodynamic behavior of trapped ion systems strongly depends on the crystal structure, resulting from the interplay between Coulomb repulsion and the trapping potential, and which can be experimentally controlled to an exquisite degree \cite{birkl92,waki92,dubin99,mavadia13,yan16}. For a highly anisotropic trapping potential the ions exhibit an inhomogeneous alignment in the axial direction \cite{ghosh95}. Due to the approximately harmonic confinement the center of the chain presents a higher density, and therefore a higher Coulomb repulsion. The decrease of the radial anisotropy can trigger structural phase transitions that start at the center of the chain and extend towards the edges as the anisotropy decreases \cite{birkl92,waki92,dubin99,mavadia13,yan16}. In particular, a second order structural phase transition from the linear chain to a planar zigzag spatial configuration has been well characterized \cite{schiffer93,piacente04,fishman08}. Also, the formation of planar concentric ellipses \cite{yan16} and three-dimensional helical configurations \cite{waki92,hasse90,nigmatullin16} has been reported.

Theoretical studies of heat transport in crystals of trapped ions connected to heat baths at different temperatures in two separate regions have revealed anomalous heat conduction, with nonlinear temperature profiles and thermal conductivities increasing with system size \cite{lin11,bermudez13,ruiz14,freitas15}. It has been shown that the linear chain exhibits almost flat temperature profiles characteristic of harmonic systems. A non-zero temperature gradient can be induced by engineered on-site disorder due to spin-vibron couplings, and  dephasing through noisy modulations of the trap frequencies \cite{bermudez13}. Also, a small amount of induced disorder using site-specific dipole forces can be used to control the transition from anomalous to normal heat transport in two- and three-dimensional crystals \cite{freitas15}. In this context, a proposal for the experimental implementation of local dephasing noise on the vibrational excitations by means of fluctuating optical potentials has been described \cite{cormick16}.    

Most of the previously mentioned studies of heat transport in trapped ion systems have considered the potential interaction in harmonic approximation about the equilibrium positions of the ions in the different spatial configurations \cite{lin11,bermudez13,freitas15}. Therefore, any possible normal heat conduction behavior had to be necessarily induced by the artificial inclusion of mechanisms such as disorder and dephasing. However, it has been shown that non-zero temperature gradients naturally arise in models that fully take into account the many-body Coulomb interaction, due to non-linearity and axial-transverse mode coupling effects arising in the proximity of the structural phase transitions \cite{ruiz14}.

The aim of this work is to study the correlation between the degree of atomic delocalization in the steady nonequilibrium dynamics of classical systems and their heat transport properties. To illustrate the analysis, we focus on a system formed by a Coulomb crystal of trapped ions, across the various structural phase transitions. We shall study a three-dimensional model corresponding to the design of a possible experiment to measure an energy current through the system. This theoretical model will be used to numerically simulate the classical dynamics of the system in contact with laser beams that emulate two heat reservoirs at different temperatures in two separate regions, until reaching the nonequilibrium steady state. Then, heat transport properties, such as temperature profiles and the total heat flux, can be obtained from the dynamical variables in such a state. We will show that the internal ions can exhibit a strong delocalization, and that such behavior is correlated with the onset of Fourier-type temperature profiles. A proper treatment of such delocalization will require a continuous description of the transport properties. 

The paper is organized as follows. In Section \ref{secII} we describe the general model considered to study the nonequilibrium dynamics and heat transport in classical systems with atomic delocalization. The nonequilibrium dynamics in the steady state is analyzed in terms of spatial probability densities, which will evidence the degree of atomic delocalization. We shall consider a continuous description to define the steady temperature profile and the total heat flux in terms of dynamical variables. In Section \ref{secIII} we particularize the model to simulate three-dimensional Coulomb crystals of trapped ions, and set the different parameters corresponding to a possible experimental setup. We show the numerical results concerning the steady state nonequilibrium dynamics of the ions for various anisotropies of the trapping potential, in particular the spatial probability densities of the entire systems and the spatial distributions of the individual ions. The temperature profiles and the total heat flux through the various structural phase transitions that modify the effective dimensionality of the trapped ion system are also shown. Finally, Section \ref{secIV} summarizes the main conclusions.   

%%%%%%%%%%%%%%%%%%%%%%%%%%%%%%%%%%%%%%%%%%%%%%%%%%%%%%%%%%%%%%%%

\section{Nonequilibrium dynamics and heat transport in classical systems with atomic delocalization}\label{secII}

%%%%%%%%%%%%%%%%%%%%%%%%%%%%%%%%%%%%%%%%%%%%%%%%%%%%%%%%%%%%%%%%

We consider a classical three-dimensional system composed of $N$ particles of mass $m$, whose motional degrees of freedom are described by the position coordinates ${\bf q}_i=(q_{x,i},q_{y,i},q_{z,i})$ and their conjugate momenta ${\bf p}_i=(p_{x,i},p_{y,i},p_{z,i})$, with $i=1,\dots,N$. We assume that the particles interact with each other through a central interaction potential $U$, and that the entire system remains confined within a finite volume due to an external trapping potential $V$. Then, the dynamics of the system can be described by the Hamiltonian
\begin{equation}
H\,=\,\sum_{i=1}^N\left[\frac{{\bf p}_i^2}{2m}+\frac{1}{2}\sum_{j\ne i}^N U_{ij}(|{\bf q}_i-{\bf q}_j|)+V_i({\bf q}_i)\right]\,.
\end{equation}

A main goal of this work is to analyze the response of this system to a imposed temperature gradient, as a function of the trapping potential $V$. The variation of such potential can be used to induce structural phase transitions that modify the effective dimensionality of the system, and therefore to control the nonequilibrium dynamics and the corresponding heat transport properties.  

\subsection{Nonequilibrium dynamics}

To induce a heat current across a given direction, we consider that $N_L$ particles on the left end along this direction and $N_R$ on the right one are connected to thermal reservoirs at different temperatures. We shall analyze a regime in which the localization induced by the interaction with the thermal reservoirs keeps these extreme particles on both ends within the spatial regions where such reservoirs are acting, whereas the remaining internal particles may be delocalized within the intermediate region. Assuming Langevin thermal reservoirs, the equations of motion for the $\alpha=(x,y,z)$ components of the position and momentum coordinates can be expressed as:
\begin{eqnarray}\label{eqm1}
&&{\dot q}_{\alpha,i}=\frac{p_{\alpha,i}}{m}
 \hspace*{0.3cm}\mbox{for}\hspace*{0.3cm} i=(1,\dots,N)\,, \\ \nonumber
&&{\dot p}_{\alpha,i}=g_{\alpha,i} + \sum_{j\ne i}^N f_\alpha^{(ij)}
-\frac{\eta_{\alpha,i}^L}{m}p_{\alpha,i}+\varepsilon_{\alpha,i}^L(t)
 \hspace*{0.3cm}\mbox{for}\hspace*{0.3cm}i=(1,\dots,N_L)\,, \\ \nonumber
&&{\dot p}_{\alpha,i}=g_{\alpha,i} + \sum_{j\ne i}^N f_\alpha^{(ij)}
 \hspace*{0.3cm}\mbox{for}\hspace*{0.3cm}i=(N_L+1,\dots,N-N_R)\,, \\ \nonumber
&&{\dot p}_{\alpha,i}=g_{\alpha,i} + \sum_{j\ne i}^N f_\alpha^{(ij)}
-\frac{\eta_{\alpha,i}^R}{m}p_{\alpha,i}+\varepsilon_{\alpha,i}^R(t)
 \hspace*{0.3cm}\mbox{for}\hspace*{0.3cm}i=(N-N_R+1,\dots,N)\,, 
\end{eqnarray}
where $g_{\alpha,i}=-\partial V_i({\bf q}_i)/\partial q_{\alpha,i}$ is the external force along the $\alpha$-direction, and $f_\alpha^{(ij)}=-f_\alpha^{(ji)}=-\partial U_{ij}(|{\bf q}_i-{\bf q}_j|)/\partial q_{\alpha,i}$ the force that the $j$th particle exerts on the $i$th particle along such direction. The action of the Langevin reservoirs is characterized by the friction coefficients $\eta_{\alpha,i}^{L,R}$ and the stochastic forces $\varepsilon_{\alpha,i}^{L,R}(t)$. This force is assumed to correspond to a Gaussian white noise that satisfies the statistical relationships
\begin{eqnarray}\label{sr}
&&\langle \varepsilon_{\alpha,i}^{L,R}(t) \rangle =0\,, \\ \nonumber
&&\langle \varepsilon_{\alpha,i}^{L,R}(t)\,\varepsilon_{\beta,j}^{L,R}(t^\prime) \rangle = 2\,D_{\alpha,i}^{L,R}\,\delta_{\alpha,\beta}\,\delta_{i,j}\,\delta(t-t^\prime)\,,
\end{eqnarray}
where $\langle\dots\rangle$ denote the average over an ensemble of stochastic trajectories, and  $D_{\alpha,i}^{L,R}$ are the diffusion coefficients. These are related to the friction coefficients $\eta_{\alpha,i}^{L,R}$ according to the fluctuation dissipation theorem \cite{kubo66}
\begin{equation}
\eta_{\alpha,i}^{L,R}=\frac{1}{2k_BT^{L,R}}\int_{-\infty}^{\infty}\,\langle\,\varepsilon_{\alpha,i}^{L,R}(t)\varepsilon_{\alpha,i}^{L,R}(t+\tau)\,\rangle\,d\tau\,=\,\frac{D_{\alpha,i}^{L,R}}{k_BT^{L,R}}\,,
\end{equation}
with $T^{L,R}$ the temperature of the corresponding thermal reservoir. 

The equations of motion (\ref{eqm1}) can be rewritten in terms of both the friction and diffusion coefficients as the following stochastic differential equations:
\begin{eqnarray}\label{eqm2}
&&dq_{\alpha,i}=\frac{p_{\alpha,i}}{m}dt
\hspace*{0.3cm}\mbox{for}\hspace*{0.3cm}i=(1,\dots,N)\,, \\ \nonumber
&&dp_{\alpha,i}=\left(g_{\alpha,i}+\sum_{j\ne i}^N f_\alpha^{(ij)}
-\frac{\eta_{\alpha,i}^L}{m}p_{\alpha,i}\right)dt+\sqrt{2\,D_{\alpha,i}^L}\,dW_{\alpha,i}^L
\hspace*{0.3cm}\mbox{for}\hspace*{0.3cm}i=(1,\dots,N_L)\,, \\ \nonumber
&&dp_{\alpha,i}=\left(g_{\alpha,i}+\sum_{j\ne i}^N f_\alpha^{(ij)}\right)dt
\hspace*{0.3cm}\mbox{for}\hspace*{0.3cm}i=(N_L+1,\dots,N-N_R)\,, \\ \nonumber
&&dp_{\alpha,i}=\left(g_{\alpha,i}+\sum_{j\ne i}^N f_\alpha^{(ij)}
-\frac{\eta_{\alpha,i}^R}{m}p_{\alpha,i}\right)dt+\sqrt{2\,D_{\alpha,i}^R}\,dW_{\alpha,i}^R
\hspace*{0.3cm}\mbox{for}\hspace*{0.3cm}i=(N-N_R+1,\dots,N)\,, 
\end{eqnarray} 
where $dW_{\alpha,i}^{L,R}$ denote the Wiener processes associated with the interactions with the laser reservoirs.

The heat transport properties of the trapped system in contact with the two thermal reservoirs are dictated by the steady state solution of the equations of motion (\ref{eqm2}). To elucidate the underlying nonequilibrium dynamics in such state we shall analyze the probability density of particles in the spatial domain ${\bf q}=(x,y,z)$. This local distribution is obtained from the set of positions that the particles have visited in their dynamics during a sufficiently long time interval $\tau_{ss}$, as:
\begin{equation}\label{rho}
P({\bf q})=\frac{1}{\tau_{ss}}\int_t^{t+\tau_{ss}}d\tau\left[\frac{1}{N\,\sigma^3(2\pi)^{3/2}}\sum_{i=1}^N\,e^{-\left[{\bf q}-{\bf q}_{i}(\tau)\right]^2/2\sigma^2}\right]\,,
\end{equation}
with $t$ an arbitrary time value within the steady state, and $\sigma$ a small parameter giving the width of the three-dimensional Gaussian kernel. The purpose of introducing the smoothing in $P({\bf q})$ is to highlight the spatial regions with high probability of finding the particles. 

In addition to the spatial probability density $P({\bf q})$ of the entire system, valuable information of the dynamics can be extracted from the analysis of the steady spatial distributions of the individual particles. In order to visualize such individual distributions, we divide the spatial region occupied by the entire system along a given $\alpha$-direction in a series composed of $c_\alpha$ cells, centered on the positions $\alpha_l\,(l=1,\dots,c_\alpha)$ and with size $\Delta$ \cite{alonso99}. During a sufficiently long time interval $\tau_{ss}$ in the nonequilibrium steady state, we monitor the passage of the particles through the various cells, and determine the time spent within each of them on successive visits. In this way, we obtain the spatial distribution of the $i$th particle along the $\alpha$-direction as:
\begin{equation}\label{theta}
\Theta_i(\alpha)\,\equiv\,\Theta_i(\alpha_l)\,=\,\frac{1}{\tau_{ss}}
\,\bigg\langle\,{\displaystyle\int_t^{t+\tau_{ss}}}d\tau\,{\displaystyle\int_{\alpha_l-\Delta/2}^{\alpha_l+\Delta/2}}\,d\alpha\,\delta\left(\,\alpha\,-\,q_{\alpha,i}(\tau)\,\right)\,\bigg\rangle\,.
\end{equation} 
In order to obtain a quasi-continuous distribution $\Theta_i(\alpha)$, a small enough value of $\Delta$ shall be considered.
   
As we will show below, strong trapping potentials $V$ lead to point like spatial distributions in which the individual particles can be clearly distinguished, whereas spatially extended distributions emerge in weaker confinements in which the internal particles can become highly delocalized.

%%%%%%%%%%%%%%%%%%%%%%%%%%%%%%%%%%%%%%%%%%%%%%%%%%%%%%%%%%%%%%%%%%%%%%%%

\subsection{Heat transport properties}

%%%%%%%%%%%%%%%%%%%%%%%%%%%%%%%%%%%%%%%%%%%%%%%%%%%%%%%%%%%%%%%%%%%%%%%%

In this section we focus on the study of the temperature profiles and the total heat flux through a selected direction, obtained from the position $\{{\bf q}\}_N=({\bf q}_1,\dots,{\bf q}_N)$ and the momentum $\{{\bf p}\}_N=({\bf p}_1,\dots,{\bf p}_N)$ coordinates of the particles in the nonequilibrium steady state reached under the action of the thermal reservoirs. We are particularly interested in analyzing the behavior of these heat transport properties through the various structural phase transitions that the system may experience due to controlled variations of the trapping potential $V$.

%%%%%%%%%%%%%%%%%%%%%%%%%%%%%%%%%%%%%%

\subsubsection{Temperature profiles}

%%%%%%%%%%%%%%%%%%%%%%%%%%%%%%%%%%%%%%

Taking into account that the internal particles may be delocalized for some configurations of the trapping potentials, we shall consider a continuous description to define the steady local temperature $T(\alpha)$ across a given $\alpha$-direction. To proceed, here again we consider a series composed of $c_\alpha$ cells along such direction, and monitor the passage of the particles through each cell during a sufficiently long time interval $\tau_{ss}$ in the nonequilibrium steady state \cite{alonso99}. We then make use of the equipartition theorem to write the temperature of the $l$th cell in terms of the kinetic energy of the particles as: 
\begin{equation}\label{temp}
T(\alpha)\,\equiv\,T(\alpha_l)=\frac{2}{3\,k_B}\,\bigg\langle\,\frac{\sum\limits_{i=1}^N{\displaystyle\int_t^{t+\tau_{ss}}}d\tau\,{\displaystyle\int_{\alpha_l-\Delta/2}^{\alpha_l+\Delta/2}}\,d\alpha\,\delta\left(\,\alpha\,-\,q_{\alpha,i}(\tau)\,\right)E_k({\bf p}_i(\tau))}{\sum\limits_{i=1}^N{\displaystyle \int_t^{t+\tau_{ss}}}d\tau\,{\displaystyle \int_{\alpha_l-\Delta/2}^{\alpha_l+\Delta/2}}\,d\alpha\,\delta\left(\alpha\,-\,q_{\alpha,i}(\tau)\,\right)}\,\bigg\rangle\,,
\end{equation} 
where $E_k({\bf p}_i)$ is the kinetic energy of the $i$th particle, and $k_B$ the Boltzmann constant.

%%%%%%%%%%%%%%%%%%%%%%%%%%%%%%%%%%%%%%

\subsubsection{Total heat flux}

%%%%%%%%%%%%%%%%%%%%%%%%%%%%%%%%%%%%%%

To obtain the heat flux across the trapped system connected to different heat reservoirs in two separate regions, we continue using a continuous description and consider the energy balance equation in local form \cite{lepri03,dhar08,zubarev74}. To proceed, we start by introducing the dynamical variable corresponding to the local energy density 
\begin{equation} 
h({\bf q})=\sum_{i=1}^N\,h_i\,\delta({\bf q}-{\bf q}_i)\,,
\end{equation}
with
\begin{equation}
h_i=\frac{{\bf p}_i^2}{2m}+V_i({\bf q}_i)+\frac{1}{2}\sum_{j\ne i}^N U_{ij}(|{\bf q}_i-{\bf q}_j|)\,.
\end{equation}
The time derivative of $h({\bf q})$, taking into account the equations of motion (\ref{eqm1}), is given by:
\begin{equation}\label{dht}
\frac{\partial h({\bf q})}{\partial t}\,=\,
\sum_{i=1}^N\left[\,\frac{1}{2}\sum_{j\ne i}^N \frac{1}{m}\left({\bf p}_i+{\bf p}_j\right)\cdot{\bf f}^{(ij)}
+J^B_i\,\right]\delta({\bf q}-{\bf q}_i)
-\nabla\cdot\sum_{i=1}^N\,\frac{{\bf p}_i}{m}\,h_i\,\delta({\bf q}-{\bf q}_i)\,,
\end{equation}
where
\begin{eqnarray}
J^B_i=\left\{
\begin{array}{cl}
\sum\limits_\alpha \frac{{p}_{\alpha,i}}{m}\left(-\frac{\eta_{\alpha,i}^L}{m}\,{p}_{\alpha,i}+\varepsilon_{\alpha,i}^L(t)\right) & \mbox{for} \hspace*{0.3cm} i=(1,\dots N_L)\,, \\
0  & \mbox{for} \hspace*{0.3cm} i=(N_L+1,\dots N-N_R)\,, \\
\sum\limits_\alpha\frac{{p}_{\alpha,i}}{m}\left(-\frac{\eta_{\alpha,i}^R}{m}\,{p}_{\alpha,i}+\varepsilon_{\alpha,i}^R(t)\right) & \mbox{for} \hspace*{0.3cm}  i=(N-N_R+1,\dots N)\,, \\
\end{array}
\right.
\end{eqnarray}
with $\alpha$ running over the components $\{x,y,z\}$, is the energy per time unit exchanged between the $i$th particle and the heat reservoir to which it is connected. 

Equation (\ref{dht}) can be rewritten as:
\begin{equation}
\frac{\partial h({\bf q})}{\partial t}\,+\,
\nabla\cdot\sum_{i=1}^N\,\frac{{\bf p}_i}{m}\,h_i\,\delta({\bf q}-{\bf q}_i)
\,-\,
\frac{1}{4}\sum_{i=1}^N\sum_{j\ne i}^N\,\frac{1}{m}\left({\bf p}_i+{\bf p}_j\right)\cdot{\bf f}^{(ij)}
\left[\,\delta({\bf q}-{\bf q}_i)-\delta({\bf q}-{\bf q}_j)\,\right]
\,=\,
\sum_{i=1}^N J^B_i\,\delta({\bf q}-{\bf q}_i)\,,
\end{equation}
where ${\bf f}^{(ij)}=-{\bf f}^{(ji)}$ has been used. To transform the last expression into a local energy balance equation, it is necessary to rewrite the third term on the right hand side in the form of a divergence. For this purpose, we take into account that 
\begin{equation}
\delta({\bf q}-{\bf q}_j)-\delta({\bf q}-{\bf q}_i)\,=\,
\nabla\cdot\int_0^1\,d\lambda\,\frac{d{\bf G}(\lambda)}{d\lambda}\,\delta({\bf q}-{\bf G}(\lambda))\,,
\end{equation} 
with ${\bf G}(\lambda)$ an arbitrary function satisfying the conditions ${\bf G}(1)={\bf q}_i$ and ${\bf G}(0)={\bf q}_j$ \cite{piccirelli68}. Then, we get the balance equation
\begin{equation}
\frac{\partial h({\bf q})}{\partial t}\,+\,\nabla\cdot{\bf j}_{h}({\bf q})=\sigma_h({\bf q})\,,
\end{equation}
with the energy flux density vector
\begin{equation}
{\bf j}_{h}({\bf q})\,=\,
\sum_{i=1}^N\,\frac{{\bf p}_i}{m}\,h_i\,\delta({\bf q}-{\bf q}_i)
\,+\,
\frac{1}{4}\sum_{i=1}^N\sum_{j\ne i}^N\,\frac{1}{m}\left({\bf p}_i+{\bf p}_j\right)\cdot{\bf f}^{(ij)}
\int_0^1\,d\lambda\,\frac{d{\bf G}(\lambda)}{d\lambda}\,\delta({\bf q}-{\bf G}(\lambda))\,,
\end{equation}
and the energy source term
\begin{equation}
\sigma_h({\bf q})\,=\,\sum_{i=1}^N J^B_i\,\delta({\bf q}-{\bf q}_i)\,.
\end{equation}
Finally, the integral of ${\bf j}_h({\bf q})$ over all space gives the total heat flux
\begin{equation}\label{Jht}
{\bf J}(t)\,=\,\sum_{i=1}^N\,\frac{{\bf p}_i}{m}\,h_i\,+\,
\frac{1}{4}\sum_{i=1}^N\sum_{j\ne i}^N\,\frac{1}{m}\left({\bf p}_i+{\bf p}_j\right)\cdot{\bf f}^{(ij)}
\,({\bf q}_i-{\bf q}_j\,)\,.
\end{equation}

In the steady state, considering that 
\begin{equation}
\big\langle\,\frac{{\bf p}_i}{m}\,h_i\,\big\rangle_{ss}\,=\,
-\big\langle\,{\bf q}_i\,\frac{dh_i}{dt}\,\big\rangle_{ss}\,=\,
-\frac{1}{2}\sum_{i=1}^N\sum_{j\ne i}^N \big\langle\frac{1}{m}\left({\bf p}_i+{\bf p}_j\right)\cdot{\bf f}^{(ij)}\,{\bf q}_i\,\big\rangle_{ss}\,-\,\sum_{i=1}^N\big\langle\,J^B_i\,{\bf q}_i\,\big\rangle_{ss}\,,
\end{equation}
where $\langle\dots\rangle_{ss}$ indicates the steady state average, the expected value of the total heat flux can be expressed as:
\begin{equation}
\langle\,{\bf J}\,\rangle_{ss}\,=\,-\sum_{i=1}^N\langle\,J^B_i\,{\bf q}_i\,\rangle_{ss}\,=\,
\frac{1}{m^2}\,\sum_i{\vphantom{\sum}}'\,\sum_\alpha\,\eta_{\alpha,i}^{B_i}\langle\,{\bf q}_i\,p^2_{\alpha,i}\,\rangle_{ss}\,-\,\frac{1}{m}\sum_i{\vphantom{\sum}}'\,\sum_\alpha\langle\,{\bf q}_i\,p_{\alpha,i}\,\varepsilon_{\alpha,i}^{B_i}(t)\,\rangle_{ss}\,.
\end{equation}
The prime denotes the sum over the particles that are connected to heat reservoirs, and $B_i\equiv L$ for $i=1,\dots,N_L$ and $B_i\equiv R$ for $i=N-N_R+1,\dots N$. According to the Novikov's theorem \cite{novikov65}, and taking into account the stochastic relationships (\ref{sr}) and the equations of motion (\ref{eqm1}), the average of the terms including the stochastic forces are given by:
\begin{equation}
\langle\,q_{\eta,i}\,p_{\alpha,i}\,\varepsilon_{\alpha,i}^{B_i}(t)\,\rangle_{ss}\,=\,
\sum_\nu\,\int\,dt^\prime\,\langle\,\varepsilon_{\alpha,i}^{B_i}(t)\,\varepsilon_{\nu,i}^{B_i}(t^\prime)\,\rangle\,\bigg\langle\,\frac{\partial(q_{\eta,i}\,p_{\alpha,i})}{\partial\varepsilon_{\nu,i}^{B_i}}\,\,\bigg\rangle_{ss}\,=\,D_{\alpha,i}^{B_i}\,q_{\eta,i}\,,
\end{equation}
with $(\eta,\alpha,\nu)$ running over the components $\{x,y,z\}$. Then, the steady total heat flux becomes
\begin{equation}\label{Jhss}
\langle\,{\bf J}\,\rangle_{ss}\,=\,\frac{1}{m^2}\,\sum_i{\vphantom{\sum}}'\sum_\alpha\,\left[\,\eta_{\alpha,i}^{B_i}\,\langle\,{\bf q}_i\,p_{\alpha,i}^2\,\rangle_{ss}\,-\,m\,D_{\alpha,i}^{B_i}\,\langle\,{\bf q}_i\,\rangle_{ss}\,\right]\,.
\end{equation}
The requirement that the ensemble average $\langle\,{\bf J}(t)\,\rangle$ of the total heat flux (\ref{Jht}) over a large enough number of stochastic trajectories coincides with the result given by (\ref{Jhss}) provides a good criterion for checking convergence to steady state in the numerical simulations.

%%%%%%%%%%%%%%%%%%%%%%%%%%%%%%%%%%%%%%

\section{Trapped ion systems}\label{secIII}

%%%%%%%%%%%%%%%%%%%%%%%%%%%%%%%%%%%%%%

In this section we deal with a three-dimensional system composed of $N$ ions of mass $m$ and charge $Q$, confined within an electromagnetic trap. We focus on the motional degrees of freedom of the ions, described by the position coordinates ${\bf q}_i=(q_{x,i},q_{y,i},q_{z,i})$ and their corresponding momenta ${\bf p}_i=(p_{x,i},p_{y,i},p_{z,i})$. We consider a system with a small number of ions, and use the pseudopotential theory to replace the trapping potential by a time-independent harmonic potential \cite{ghosh95}. We assume that this approximation, which neglects the rapid micromotion, captures the essence of the secular motion. Specifically, we assume that the ions are confined by a harmonic trap with the axial frequency $\omega_x$, and the transverse (radial) frequencies $\omega_y$ and $\omega_z\,$. To characterize the radial anisotropy of the trap we introduce the parameters $n_\beta=\omega_\beta/\omega_x$, with $\beta=(y,z)$. Then, the dynamics of the system is ruled by both the interaction with the harmonic trap 
\begin{equation}
V_i({\bf q}_i)=\frac{1}{2}\,m\,\omega_x^2\,(\,q_{x,i}^2+n_y^2\,q_{y,i}^2+n_z^2\,q_{z,i}^2\,)\,,
\end{equation}
and the Coulomb repulsion
\begin{equation}
U_{ij}(|{\bf q}_i-{\bf q}_j|)\,=\,\left(\frac{Q^2}{4\pi\epsilon_0}\right)\,\frac{1}{|{\bf q}_i-{\bf q}_j|}\,,
\end{equation}
with $\epsilon_0$ the vacuum permittivity.

In this trapped ion system we study the steady state nonequilibrium dynamics as well as heat transport, through the various structural phase transitions induced by the variation of anisotropy of the trapping potential. We focus on the study of the spatial distributions of the ions, and the temperature profiles and the total heat flux across the axial direction. To induce a heat current across the system, we consider that the $N_L$ leftmost ions along the $x$-direction and the $N_R$ rightmost ones are connected to laser beams that simulate two thermal reservoirs at different temperatures. In order to resolve the dynamics we assume that such laser beams can be modeled as Langevin heat baths. In addition, considering that the typical separations between the ions are generally of the order of micrometers, we adopt a classical description of the nonequilibrium dynamics. Then, such dynamics can be described by the equations of motion (\ref{eqm2}).

For small laser intensities, the friction coefficients $\eta_{\alpha,i}^{L,R}$ and the diffusion coefficients $D_{\alpha,i}^{L,R}$ can be obtained from the Doppler cooling expressions \cite{phillips92}:
\begin{equation}
\eta_{\alpha,i}^{L(R)}=-4\,\hbar\,\left(k_{\alpha,i}^{L(R)}\right)^2\left(\frac{I_{\alpha,i}^{L(R)}}{I_0^{L(R)}}\right)\,\frac{\left(2\,\delta_{\alpha,i}^{L(R)}/\Gamma\right)}{\left[1+4\left(\delta_{\alpha,i}^{L(R)}\right)^2\,/\,\Gamma^2\right]^2}
\end{equation}
and
\begin{equation}
D_{\alpha,i}^{L(R)}=\hbar^2\,\left(k_{\alpha,i}^{L(R)}\right)^2\left(\frac{I_{\alpha,i}^{L(R)}}{I_0^{L(R)}}\right)\,\frac{\Gamma}{\left[1+4\left(\delta_{\alpha,i}^{L(R)}\right)^2\,/\,\Gamma^2\right]}\,.
\end{equation}
The ratio $I_{\alpha,i}^{L(R)}/I_0^{L(R)}$ denotes the normalized intensity of the laser beam acting on the $i$th ion along the $\alpha$-direction, $k^{L(R)}_{\alpha,i}$ is the corresponding laser wavelength, $\delta_{\alpha,i}^{L(R)}=\omega_{\alpha,i}^{L(R)}-\omega_0$ is the detuning of the laser frequency $\omega_{\alpha,i}^{L(R)}$ with respect to the frequency $\omega_0$ of a selected atomic transition in the ion, and $\Gamma$ is the natural linewidth of the excited state in such transition. In this work we select the atomic transition $3s^2S_{1/2}\rightarrow3p^2P_{1/2}$ of the $^{24}$Mg$^+$ ions, with $\omega_0/2\pi=1069$ THz and $\Gamma/2\pi=41.296$ MHz \cite{nist99}. To induce a heat current through the trapped ion system we shall consider that the extreme ions on both ends are subjected to laser beams with different detunings $\delta_{\alpha,i}^{L}\ne\delta_{\alpha,i}^{R}$. From now on we set $\delta_{\alpha,i}^{L}=-0.02\Gamma$ for the $N_L$ leftmost ions and $\delta_{\alpha,i}^{R}=-0.1\Gamma$ for the $N_R$ rightmost ones, and the same laser intensity $I_{\alpha,i}^{L(R)}/I_0=0.08$ on both ends. These values lead to the friction coefficients $\eta^L=6.76\times 10^{-22}\,$kg$/$s and $\eta^R=3.13\times 10^{-21}\,$kg$/$s, and the diffusion coefficients $D^L=1.16\times 10^{-46}\,$kg$^2$m$^2$$/$s$^3$ and $D^R=1.11\times 10^{-46}\,$kg$^2$m$^2$$/$s$^3$. For reference, the corresponding limit Doppler temperatures, obtained in the case of the Doppler cooling of a single isolated ion, would be $T^L=D^L/k_B\eta^L=12.41\,$mK and $T^R=D^R/k_B\eta^R=2.58\,$mK. Thus the trapped ion system shall be connected to an effective hotter bath on the left end, and to a colder bath on the right end. 

%friction coeff. (L) (kg/s):   6.7584421979003376E-022   6.7584421979003376E-022   6.7584421979003376E-022
%diffusion coeff. (L) (kg^2m^2/s^3):  1.1576703640156898E-046   1.1576703640156898E-046   1.1576703640156898E-046
%T_L =   12.406666037783301     
%friction coeff. (R): 3.1342855365440246E-021   3.1342855365440246E-021   3.1342855365440246E-021
%diffusion coeff. (R): 1.1149256052245360E-046   1.1149256052245360E-046   1.1149256052245360E-046
%T_R (mK): 2.5764641931498864     

%%%%%%%%%%%%%%%%%%%%%%%%%%%%%%%%%%%%%%%%%%%%%%%%%%%%%%%%%%%%%%%%

\subsection{Spatial configurations of the trapped ions}

%%%%%%%%%%%%%%%%%%%%%%%%%%%%%%%%%%%%%%%%%%%%%%%%%%%%%%%%%%%%%%%%

Before dealing with the study of the heat transport properties, we analyze the underlying nonequilibrium dynamics of the trapped ions according to the anisotropy of the confining potential. In this section we present a detailed analysis of the steady state spatial distribution of the trapped ions as a function of the parameters of the radial anisotropy $(n_y,n_z)$. We are particularly interested in identifying the values of these parameters leading to the different structural phase transitions. 

To perform the numerical analysis, from now on we shall consider a system composed of $N=30$ $^{24}$Mg$^+$ ions, with $N_L=N_R=3$ of them connected to the laser reservoirs on both ends along the axial direction. We set the axial frequency $\omega_x/2\pi=50\,$kHz, and study the dynamics for different transverse frequencies $(\omega_y,\omega_z)$. In order to ensure that the extreme ions that are connected to the lasers reservoirs remain within the space region where the beams are focussed, the radial frequencies must be sufficiently high. For a fixed value of $\omega_x$, this imposes a lower limit to the values of $(n_y,n_z)$. In the case of the selected value of $\omega_x$, this requires considering values of both anisotropy parameters above approximately 6.7.

%%%%%%%%%%%%%%%%%%%%%%%%%%%%%%%%%%%%%%%%%%%%%%%%%%%%%%%%%%%%%%%%%%
\begin{figure}[h]\centering
\includegraphics[width=0.55\linewidth]{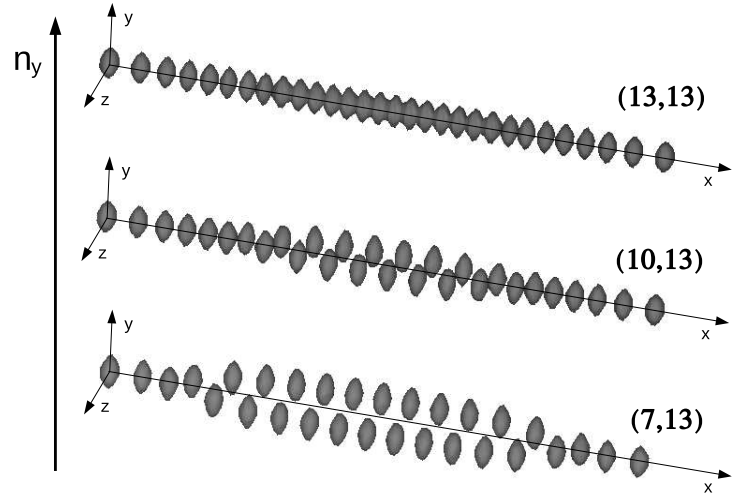}
\caption{
Spatial probability density $P({\bf q})$ (\ref{rho}) of the trapped ion system, obtained from a single stochastic realization, with $t\,=\,\tau_{ss}\,=\,5\times 10^{-2}\,s$ and $\sigma\,=\,2\mu m$.  The axial frequency is set to $\omega_x/2\pi=50\,$kHz, the radial frequency $\omega_z=13\omega_x$ and each configuration corresponds to a different radial frequency $\omega_y=n_y\omega_x$. The labels are the values of the anisotropy parameters $(n_y,n_z)$. The positions ${\bf q}$ with values of $P({\bf q})$ below $5$\% of its maximum value are not depicted. The total extension of the system along the axial direction is reduced from approximately $420\,\mu m$ in the linear chain to $402\,\mu m$ in the complete zigzag configuration. The VESTA software was used for the visualization of the spatial distributions \cite{vesta}.
}
\label{nz13a}
\end{figure}
%%%%%%%%%%%%%%%%%%%%%%%%%%%%%%%%%%%%%%%%%%%%%%%%%
%%%%%%%%%%%%%%%%%%%%%%%%%%%%%%%%%%%%%%%%%%%%%%%%%
\begin{figure}[h]\centering
\includegraphics[width=0.55\linewidth]{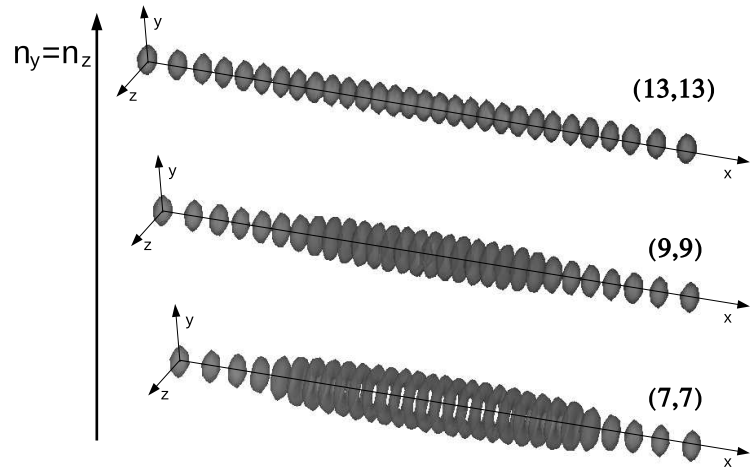}
\caption{
The same as Fig. \ref{nz13a}, for fixed axial frequency $\omega_x/2\pi=50\,$kHz and three pairs of equal radial frequencies $\omega_y=\omega_z$. In this case, the positions ${\bf q}$ with values of $P({\bf q})$ below $6.5$\% of its maximum value are not depicted. The total extension of the system along the axial direction is reduced from approximately $420\,\mu m$ in the linear chain to $402\,\mu m$ in the complete helical configuration. 
}
\label{nynza}
\end{figure}
%%%%%%%%%%%%%%%%%%%%%%%%%%%%%%%%%%%%%%%%%%%%%%%%%
%%%%%%%%%%%%%%%%%%%%%%%%%%%%%%%%%%%%%%%%%%%%%%%%%
\begin{figure}[h]\centering
\includegraphics[width=0.55\linewidth]{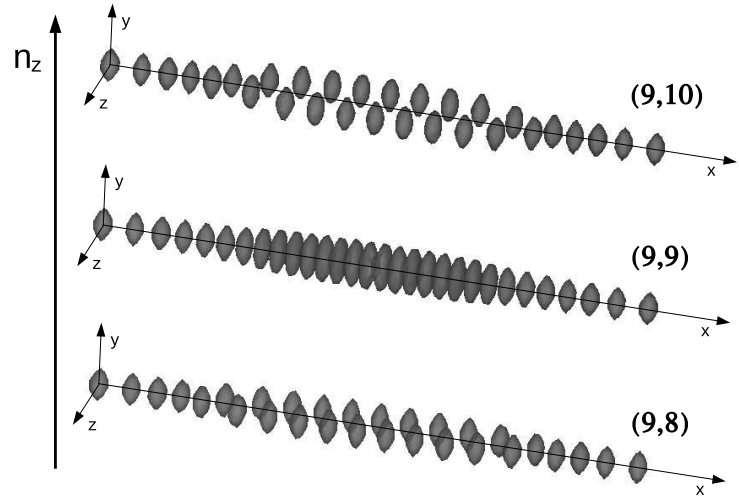}
\caption{
The same as Fig. \ref{nz13a}, for fixed axial frequency $\omega_x/2\pi=50\,$kHz, radial frequency $\omega_y=9\omega_x$ and three different radial frequencies $\omega_z=n_z\omega_x$. The total extension of the system along the axial direction is approximately $410\,\mu m$ in the three configurations. Notice that in the lowest panel the planar zigzag configuration is confined in the $xz$-plane, whereas in the upper one is in the $xy$-plane. 
}
\label{ny9a}
\end{figure}
%%%%%%%%%%%%%%%%%%%%%%%%%%%%%%%%%%%%%%%%%%%%%%%%%

In the numerical simulations we set the initial conditions with the ions at rest, and positions randomly distributed in the close vicinity of the equilibrium positions of the linear configuration along the axial direction. Then the laser reservoirs that are focused on the selected ions are switched on instantaneously. We perform the time evolution until the system reaches the nonequilibrium steady state from which the energy transport properties are extracted. Figures \ref{nz13a}, \ref{nynza} and \ref{ny9a} show the steady state spatial probability densities (\ref{rho}) of the ions for different anisotropies of the trapping potential.

Figure \ref{nz13a} illustrates the change of the spatial probability distribution through a linear-zigzag structural phase transition. It shows how for a fixed axial frequency $\omega_x$ such transition spreads from the center to the edges as the radial anisotropy $n_y$ is lowered, whereas the high radial anisotropy $n_z$ confines the ions in the $xy$-plane. 

As Fig. \ref{nynza} shows, in the case of a symmetrical trap with equal radial frequencies, their decrease gives rise to a transition from the linear configuration to a three-dimensional one, in which the ions distribute over a series of rings contained in the transverse $yz$-plane and centered along the axial direction $x$. As occurs with the zigzag configuration, the rings arise at the center of the system and extend towards the ends as the radial frequencies decrease. This three-dimensional configuration of the trapped ions corresponds to the helical arrangement reported in previous studies \cite{waki92,hasse90,nigmatullin16}. From now on we will refer to it as the helical configuration.

Outside the high frequency domain corresponding to the linear configuration, the variation of one of the transverse frequencies through the helical configuration $(n_y=n_z)$ results in a rotation of the zigzag configuration around the axial axis, to be confined again in the plane perpendicular to transverse direction with the highest anisotropy. As an illustration, Fig. \ref{ny9a} shows the transition from the zigzag configuration confined in the $xz$-plane, given by $(n_y,n_z)=(9,8)$, to the zigzag configuration in the $xy$-plane, given by $(n_y,n_z)=(9,10)$, through the helical configuration corresponding to $(n_y,n_z)=(9,9)$.

\begin{figure}[h]\centering
\includegraphics[width=0.53\linewidth]{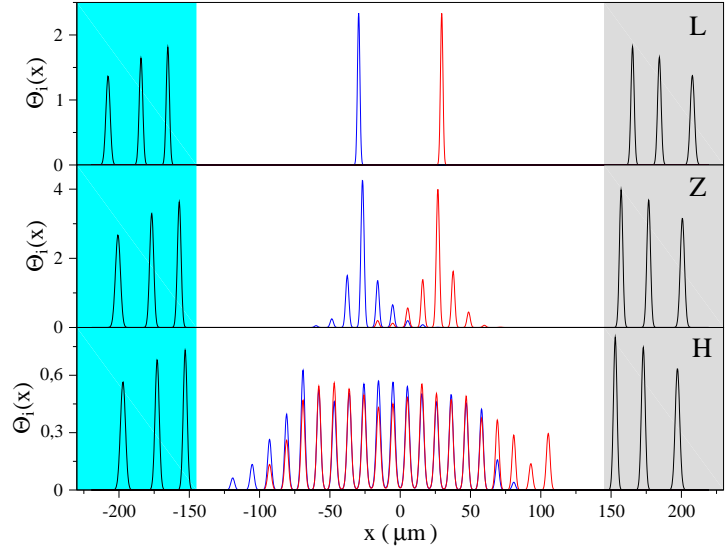}
\caption{Steady axial distributions $\Theta_i(x)$ (\ref{theta}) for the same eight ions in the linear (L) configuration given by $(n_y,n_z)=(13,13)$, in the zigzag (Z) configuration given by $(n_y,n_z)=(9,8)$, and in the helical (H) configuration given by $(n_y,n_z)=(7,7)$.  We consider $c_x=2000$ cells, with size $\Delta=0.22\mu m$, along the $x$-direction, set the time values $t\,=\,\tau_{ss}\,=\,4\times 10^{-2}\,s$, and average over more than $500$ stochastic trajectories. The peaks on both sides (black line) correspond to the $N_L=3$ and $N_R=3$ extreme ions that are connected to the laser reservoirs, and evidence the strong spatial confinement of such ions in the spatial regions where the beams are focussed. Such regions are depicted by the colored areas. The blue and red lines correspond to the same two internal ions. For a better comparison, the intensity of the six peaks of the extreme ions (black line) has been divided by a factor of two in the middle panel and by a factor of eleven in the lower one. 
}
\label{x_sd}
\end{figure}

\begin{figure}[h]\centering
\vspace*{0.3cm}
\includegraphics[width=0.53\linewidth]{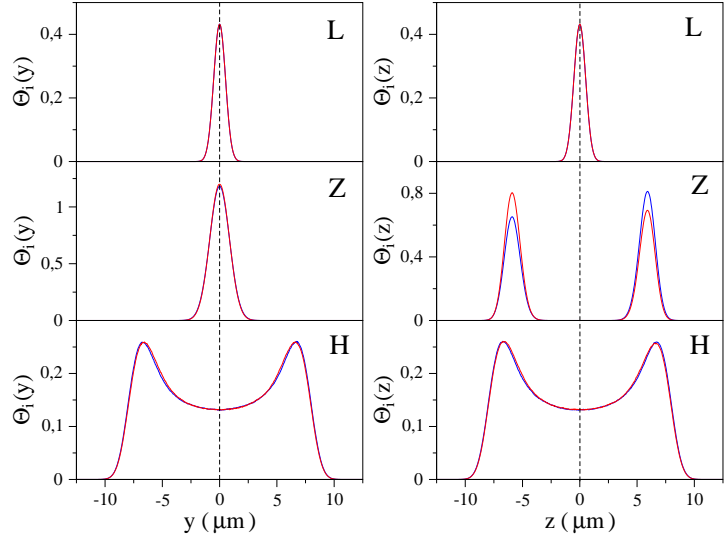}
\caption{Steady radial distributions $\Theta_i(y)$ and $\Theta_i(z)$ (\ref{theta}) of the two internal ions (blue and red lines) whose axial distributions $\Theta_i(x)$ are depicted in Fig. \ref{x_sd}, for the linear (L), zigzag (Z) and helical (H) configurations considered in such figure. We consider $c_y=c_z=2000$ cells, with size $\Delta=0.024\mu m$, along both radial directions, and set the time values and perform the stochastic average as in Fig. \ref{x_sd}.
}
\label{yz_sd}
\end{figure}

So far we have used the spatial probability densities of the entire trapped ion system to illustrate the different structural phase transitions that it may experience. Notice that while each of the $N_L=N_R=3$ outermost dots, as well as all dots in the linear chain, can be assigned to specific ions, we will now show that this not necessarily the case for the internal dots of the zigzag configuration, nor for the various rings of the helical configuration. To elucidate the dynamics performed by the individual ions in the three previously shown spatial configurations, we now analyze the steady spatial distributions (\ref{theta}). Figure \ref{x_sd} illustrates some of such individual distributions along the axial direction, and Fig. \ref{yz_sd} along the two transversal directions.  

%%%%%%%%%%%%%%%%%%%%%%%%%%%%%%%%%%%%%%%%%%%%%%%%%
%%%%%%%%%%%%%%%%%%%%%%%%%%%%%%%%%%%%%%%%%%%%%%%%%
%%%%%%%%%%%%%%%%%%%%%%%%%%%%%%%%%%%%%%%%%%%%%%%%%
%%%%%%%%%%%%%%%%%%%%%%%%%%%%%%%%%%%%%%%%%%%%%%%%%

As expected, in the linear chain $\Theta_i(\alpha)$ exhibits a single peak centered at the corresponding equilibrium position along the axial axis, for all ions $i=1,\dots N$ and directions $\alpha=(x,y,z)$, see Figs. \ref{x_sd} and \ref{yz_sd}. In this configuration each ion is strongly confined, and can only perform small oscillations around its equilibrium position. 

In the zigzag and helical configurations the distributions $\Theta_i(x)$ of the internal ions, see Fig. \ref{x_sd}, exhibit a series of peaks, which evidence the delocalization of these ions along the axial direction. The intensities of the various peaks vary from one ion to another, but their positions are the same, as they are determined by the minima of the global potential energy surface. In the zigzag configuration the different peaks correspond to equilibrium positions located on both sides of the axial axis, see for example Fig. \ref{nz13a}. Whereas in the helical configuration they give the location of the series of rings shown in Fig. \ref{nynza}. Thus, the lower confinement of the internal ions allows them to move throughout the system and exchange their positions along the axial direction. This axial displacement can occur along practically all the region covered by the helical configuration, through very fast jumps between the different rings. In the zigzag configuration such displacement is more local, as it is restricted to rapid jumps between neighboring equilibrium positions.

The radial distributions $\Theta_i(y)$ and $\Theta_i(z)$ clearly distinguish between the linear, zigzag and helical configurations, see Fig. \ref{yz_sd}. In the zigzag configuration the distribution of the radial coordinate corresponding to the highest trapping frequency presents a single peak centered at zero, while the other radial coordinate exhibits two peaks of similar intensity arranged symmetrically around zero. The probability of presence practically zero in between the two peaks indicates that the ions are jumping very rapidly through the axial axis, staying most of the time in the close vicinity of any of the minima that the potential has on either side of this axis. In the helical configuration both radial coordinates show nearly identical bimodal distributions, again arranged symmetrically around zero. But in this case there is a significant probability of presence in between both maxima, which evidences the distribution of the ions within the rings shown in Fig. \ref{nynza}.

\begin{figure}[t]\centering
\includegraphics[width=0.6\linewidth]{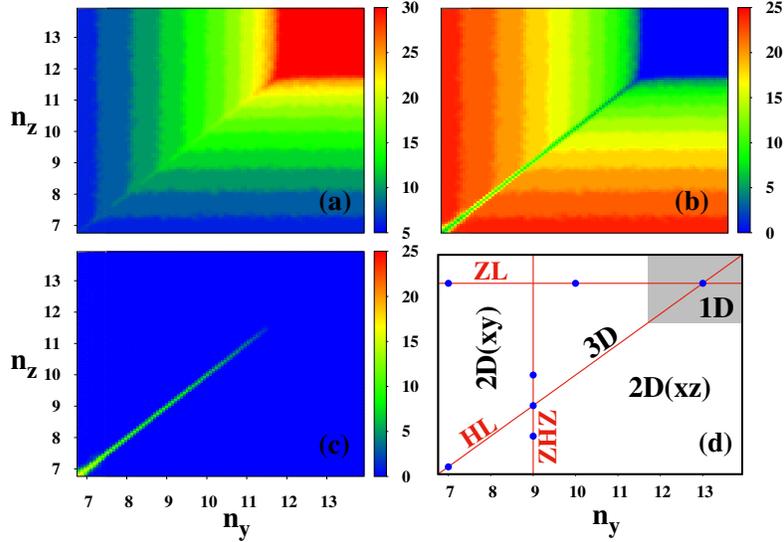}
\caption{The number of ions arranged in the linear (a), the zigzag (b), and the helical configuration (c), obtained from the numerical simulation of the dynamics in the interval $[1\times 10^{-2},2\times 10^{-2}]\,s$ of the steady state. A diagram (d) of the anisotropy map that shows the values of $(n_y,n_z)$ leading to the linear string along the axial axis (gray square area at high values of both $n_y$ and $n_z$), the zigzag configuration in the $xz$-plane (below the diagonal $n_y=n_z$), the zigzag configuration in the $xy$-plane (above the diagonal $n_y=n_z$), and the helical configuration (along the diagonal $n_y=n_z$). In panel (d), the values of anisotropy parameters $(n_y,n_z)$ corresponding to the spatial probability densities depicted in Figs. \ref{nz13a}, \ref{nynza} and \ref{ny9a} are shown (blue circles). The ZL- (zigzag-linear transition), HL- (helical-linear transition) and ZHZ-line (zigzag-helical-zigzag transition) considered to perform the analysis of the transport properties are also indicated. 
}
\label{fm}
\end{figure}

Taking into account the delocalization that the internal ions can exhibit, we shall characterize the configuration changes using an alternative order parameter to those considered in previous studies \cite{schiffer93,freitas15,yan16}. Concretely, we have shown that the two radial distributions $\Theta_i(y)$ and $\Theta_i(z)$ are highly sensitive to changes in the global potential energy surface leading to the structural phase transitions. We now employ the locations of their maxima as a criterion to construct an anisotropy map that shows the values of $(n_y,n_z)$ at which the different spatial configurations occur. To be more precise, the maxima of both distributions are located at zero for an ion aligned with the chain axis, only one of the two distributions has the maximum at zero for an ion that is in a zigzag configuration, whereas neither of them presents a maximum at zero for an ion arranged in a helical configuration. The anisotropy map obtained for the ion distributions in the three different spatial configurations is given in Fig. \ref{fm}.

%%%%%%%%%%%%%%%%%%%%%%%%%%%%%%%%%%%%%%%%%%%%%%%%%%
%%%%%%%%%%%%%%%%%%%%%%%%%%%%%%%%%%%%%%%%%%%%%%%%%%%%%%%%%%%%%%%%%
As expected, all the ions are located along the axial direction in trapping potentials with high radial anisotropy. In the case of a chain composed of $N=30$ $^{24}$Mg$^+$ ions, we observe that this occurs for values of both $n_y$ and $n_z$ above approximately $11.6$, see Fig. \ref{fm}. Outside such region the trapped ion system exhibits predominantly zigzag configurations in the plane perpendicular to the radial direction with the largest anisotropy, and with a decreasing number of external ions along the axial axis as the radial trapping frequencies become smaller. The helical configuration does not occur unless the two radial trapping frequencies become practically equal. Thus, it emerges as a distinctive feature of traps with symmetrical anisotropy, where $\omega_y=\omega_z\,$, in contrast to the ubiquitous zigzag configurations for radially asymmetric traps, in which $\omega_y\ne\omega_z$, provided at least one of the two radial frequencies is sufficiently small.

%%%%%%%%%%%%%%%%%%%%%%%%%%%%%%%%%%%%%%%%%%%%%%%%%%%%%%%%%%%%%%%%%%%%%%%%

\subsection{Temperature profiles and total heat flux}

%%%%%%%%%%%%%%%%%%%%%%%%%%%%%%%%%%%%%%%%%%%%%%%%%%%%%%%%%%%%%%%%%%%%%%%%

Now we analyze the steady state temperature profiles and the total heat flux across the axial direction, for various anisotropies of the trapping potential. In particular, we focus on their behavior through the various structural phase transitions shown in the previous section.
%%%%%%%%%%%%%%%%%%%%%%%%%%%%%%%%%%%%%%%%%%%%%%%%%
\begin{figure}[h]\centering
\includegraphics[width=0.55\linewidth]{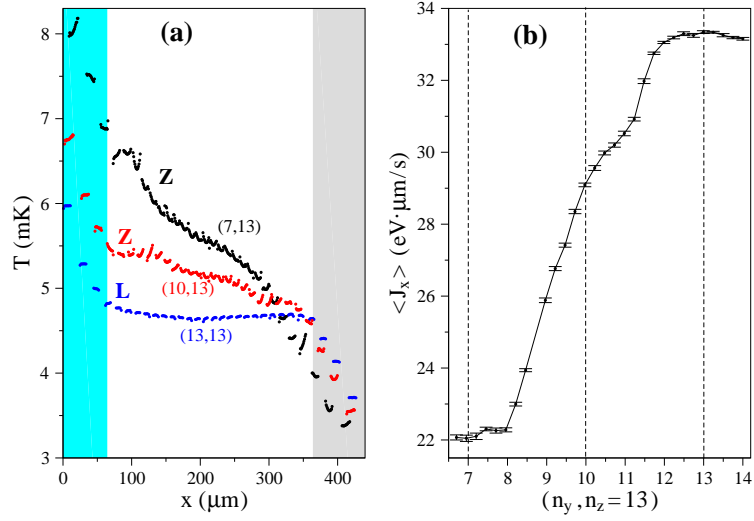}
\caption{(a) Steady state temperature profiles $T(x)$ (\ref{temp}) across the axial direction for the linear (L) and the zigzag (Z) configurations depicted in Fig. \ref{nz13a}. The colored areas are the regions where the two heat reservoirs are acting. (b)  The component of the steady total heat flux ${\bf J}$ (\ref{Jht}) across the axial direction for anisotropy parameters selected along the ZL-line depicted in Fig. \ref{fm}, corresponding to the zigzag-linear structural phase transition occurring for $n_z=13$ and different values of $n_y$. The dashed lines indicate the anisotropy parameters whose temperatures profiles are shown in (a). The results were obtained from numerical simulations of the dynamics in the interval $[4\times 10^{-2},8\times 10^{-2}]\,s$ of the steady state, and the average over more than $1000$ stochastic trajectories. The error bars in the total heat flux provide a measure of the fluctuations around such average, and are given by the corresponding standard deviations. We consider $c_x=500$ cells, with size $\Delta=0.85\mu m$, along the axial direction to get the temperature profiles.  
}
\label{fig_zl}
\end{figure}
%%%%%%%%%%%%%%%%%%%%%%%%%%%%%%%%%%%%%%%%%%%%%%%%%
%%%%%%%%%%%%%%%%%%%%%%%%%%%%%%%%%%%%%%%%%%%%%%%%%
\begin{figure}[h]\centering
\vspace*{0.6cm}
\includegraphics[width=0.55\linewidth]{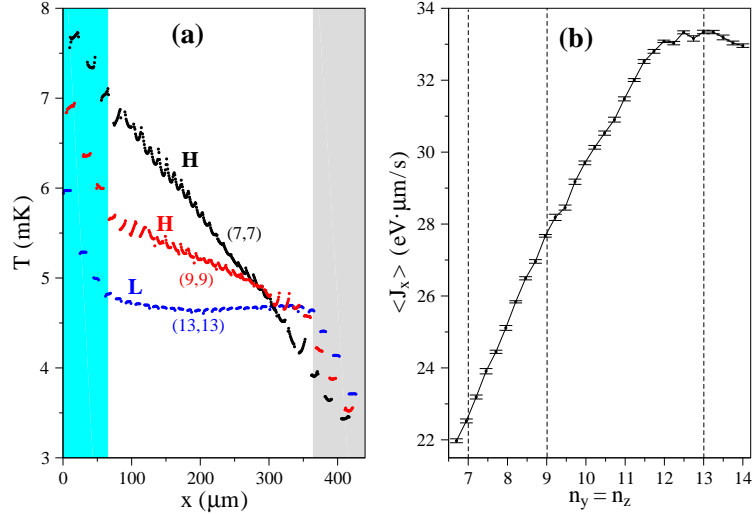}
\caption{The same as Fig. \ref{fig_zl}; (a) for the linear (L) and the helical (H) configurations depicted in Fig. \ref{nynza}, and (b) for anisotropy parameters selected along the HL-line depicted in Fig. \ref{fm}, corresponding to the helical-linear structural phase transition occurring for different values of $n_y=n_z$.     
}
\label{fig_hl}
\end{figure}
%%%%%%%%%%%%%%%%%%%%%%%%%%%%%%%%%%%%%%%%%%%%%%%%%
%%%%%%%%%%%%%%%%%%%%%%%%%%%%%%%%%%%%%%%%%%%%%%%%%
\begin{figure}[h]\centering
\includegraphics[width=0.55\linewidth]{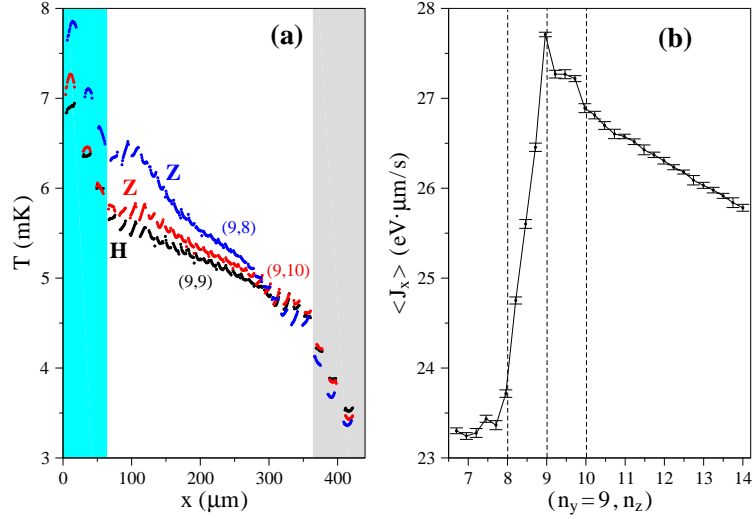}
\caption{The same as Fig. \ref{fig_zl}; (a) for the zigzag (Z) and the helical (H) configurations depicted in Fig. \ref{ny9a}, and (b) for anisotropy parameters selected along the ZHZ-line depicted in Fig. \ref{fm}, corresponding to the zigzag-helical-zigzag structural phase transition occurring for $n_y=9$ and different values of $n_z$.
}
\label{fig_zhz}
\end{figure}
%%%%%%%%%%%%%%%%%%%%%%%%%%%%%%%%%%%%%%%%%%%%%%%%%
Figure \ref{fig_zl}(a) shows the temperature profile and the total heat flux across a linear-zigzag structural phase transition, Fig. \ref{fig_hl}(a) in a linear-helical transition, and Fig. \ref{fig_zhz}(a) in a transition between two perpendicular zigzag configurations through a helical configuration. 

The presence of separate segments across the various temperature fields is due to the strong spatial confinement of certain ions around their equilibrium positions. In agreement with the results presented in the previous section, such confinement persists for all ions in the linear string, while in the zigzag and helical configurations the delocalization of the internal ions leads to quasi-continuous central regions in the temperature profiles.

The small size of the system leads to significant boundary effects in the temperature profiles, mainly in the regions occupied by the ions that are connected to the thermal baths and their nearest neighbors. In between those regions, the temperature field corresponding to the internal ions becomes very sensitive to the structural phase changes. In the linear chain it shows a nearly perfect plateau.  As is known, a flat temperature profile evidences the anomalous heat transport in one-dimensional harmonic crystals, which have infinite conductivity and can not, therefore, support a temperature gradient \cite{rieder67,dhar16}. 

The structural phase transitions from the linear to the zigzag configuration and from the linear to the helical configuration introduce a non-zero temperature gradient with the axial coordinate, whose magnitude increases as the transitions spread from the center to the edges. While in the zigzag configuration the overall gradient is not uniform, in the helical one it remains almost constant, giving rise to a quasi-linear temperature profile consistent with Fourier's law.

These results elucidate a correlation between the amount of delocalization of the internal ions and the temperature profiles across the axial direction. While an almost complete delocalization in the helical configuration leads to a Fourier like behavior, the strong confinement in the linear chain results in the nearly flat profiles characteristic of an anomalous heat transport in harmonic systems. The more restricted delocalization in the zigzag configuration corresponds to an intermediate behavior, with non fully linear temperature profiles. 

%%%%%%%%%%%%%%%%%%%%%%%%%%%%%%%%%%%%%%%%%%%%%%%%%
\begin{figure}[h]\centering
\vspace*{0.5cm}
\includegraphics[width=0.55\linewidth]{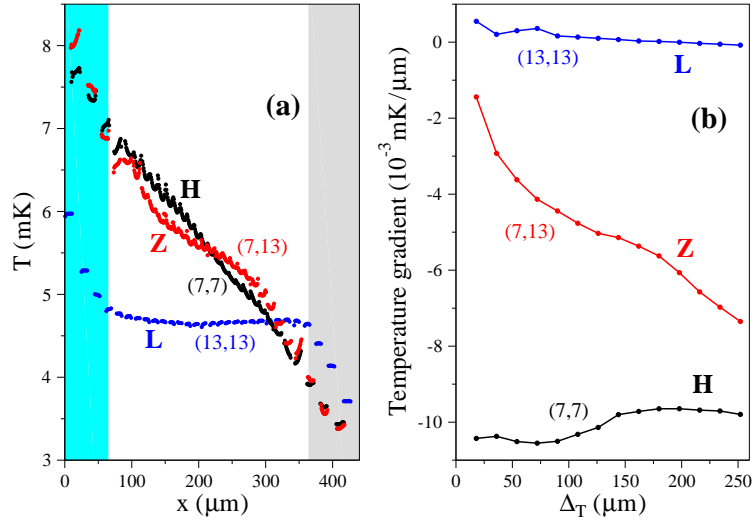}
\caption{(a) Steady state temperature profiles $T(x)$ (\ref{temp}) across the axial direction for the complete linear (L), zigzag (Z) and helical (H) configurations. The colored areas are the regions where the two heat reservoirs are acting. The labels indicate the values of the anisotropy parameters $(n_y,n_z)$. The ion spatial distributions corresponding the different configurations are depicted in Figs. \ref{nz13a} (L and Z) and \ref{nynza} (L and H). (b) Temperature gradients across the region occupied by the internal ions, corresponding to the temperature profiles shown in (a), and obtained for intervals of the $x$-coordinate symmetrically arranged around the center of the system and with increasing size $\Delta_T$.
}
\label{fig_zhl}
\end{figure}
%%%%%%%%%%%%%%%%%%%%%%%%%%%%%%%%%%%%%%%%%%%%%%%%%

As Fig. \ref{fig_zhl}(a) summarizes, the temperature profiles corresponding to the complete linear, zigzag and helical configurations can be clearly differentiated from each other. The analysis of the temperature gradient across the axial region occupied by the internal ions clearly shows the different behaviors of the three configurations, see Fig. \ref{fig_zhl}(b). As expected, this gradient remains very close to zero in the linear chain. In the helical configuration it presents very small variations around a non-zero value, which is consistent with Fourier's law. While the zigzag configuration clearly departs from the behavior predicted by this law, by exhibiting a significant variation of the temperature gradient.

The temperature profiles obtained for the linear and zigzag configurations are in agreement with those previously reported using a discrete description in a two-dimensional model \cite{ruiz14}, as in such configurations the delocalization of the internal ions is absent or remains restricted. However, as shown in this work, a proper analysis of the helical configuration necessarily requires a continuous description that takes into account the displacement of the internal ions across the axial direction. 

The spectral analysis of the steady state evolution of the coordinates of the internal ions has shown that the formation of a non-zero temperature gradient, as the system experiences a structural phase transition from the linear chain to configurations of higher dimensionality, can be assigned to non-linearities and the coupling between radial and axial modes, which are absent in the harmonic picture \cite{ruiz14}. Such effects become also visible in the total heat flux across the axial direction, which exhibits signatures of the structural phase transitions, see Figs. \ref{fig_zl}(b), \ref{fig_hl}(b) and \ref{fig_zhz}(b). 

On approaching the transitions from the high anisotropy domain corresponding to the linear chain, the increasing contribution of the transverse modes and the growing level of fluctuations due to thermal motion of the ions may assist transport, leading to a progressive increase of the heat flux \cite{ruiz14}. As Fig. (\ref{fig_zl})(b) and (\ref{fig_hl})(b) show, once the transition has already emerged and the chain buckles, a further decrease of the radial anisotropy leads to a reduction of the total heat flux as the transition spreads from the center to the edges and the internal ions jump off the axial axis.

The reduction of the heat flux can be attributed to the decreasing interaction between neighboring ions as they arrange in configurations of higher dimensionality, in which the inter-ion distances become larger. While in the linear-helical transition such reduction is nearly uniform, in the linear-zigzag transition it tends to stabilize at low anisotropies, once all the internal ions have jumped off the axial axis. In the case of a system composed of $N=30$ $^{24}$Mg$^+$ ions, the total reduction of the heat flux through the transition from the linear chain to the complete zigzag and helical configurations reaches around 34\%.

The analysis of the linear-zigzag and linear-helical transitions has shown that heat transport is optimal in the linear configuration, in the proximity of the onset of the structural phase transition. According to Fig. \ref{fig_zhz}(b), in the case of the transition between two perpendicular zigzag configurations, the heat flux exhibits a maximum at the intermediate helical configuration. Thus the coupling between the axial and transverse modes becomes more detrimental to heat transport in the zigzag configuration than in the helical configuration with a similar number of internal ions located outside the axial axis.

%%%%%%%%%%%%%%%%%%%%%%%%%%%%%%%%%%%%%%%%%%%%

\section{Conclusions}\label{secIV}

We have used Coulomb crystals of trapped ions to get deeper insight into the connection between heat transport properties and the underlying nonequilibrium dynamics in systems that are in contact with different thermal baths in two separate regions. We have considered an intrinsically non-linear three dimensional model, which fully takes into account the many-body Coulomb interaction, and analyzed the response of the system through the various structural phase transitions induced by the controlled variation of the anisotropy of the trapping potential. 

The results, obtained from the numerical resolution of the classical Langevin equations of motion, have shown a correlation between the degree of delocalization of the internal ions and the temperature gradient across the axial direction. The strong confinement of the ions around their equilibrium positions in the linear chain leads to nearly flat temperature profiles characteristic of the anomalous heat conduction one-dimensional harmonic systems. While the extended delocalization of the ions in the helical configuration is associated with global quasi-linear temperature profiles, as predicted by Fourier's law. The planar zigzag configuration corresponds to an intermediate situation, in which the more restricted delocalization results in non-zero, but non-uniform temperature gradients.

Considering that Fourier's law is a macroscopic consequence of ordinary diffusion at the microscopic level, the observed delocalization of the internal ions across the entire helical configuration could correspond to a ordinary Brownian diffusion. This is not expected in the case of the zigzag configuration, for which the temperature profiles do not fully exhibit a Fourier-type behavior.

While previous studies based on harmonic models have shown that the onset of nonzero temperature gradients across the trapped ion system can be artificially induced by engineered dephasing and disorder, in this work we show that the transition from anomalous to normal transport arises naturally in an intrinsically non-linear model. The interplay between the many-body Coulomb interaction and the external substrate potential of the trap leads to a very rich underlying nonequilibrium dynamics, which is ultimately responsible for the strong dependence of the transport properties on the structural phase transitions that modify the dimensionality of the system.  

We have shown that the total heat flux across the axial direction is highly sensitive to changes in the effective dimensionality of the trapped ion system. Heat transport is optimal in the linear configuration, in the proximity of the onset for the structural phase transitions to configurations of larger dimensionality. This may be attributed to an increasing contribution of the transverse modes to transport, and the increasing thermal motion of the ions. Upon further decrease of the anisotropy of the trapping potential, the spread of the structural phase transitions across the axial direction results in a progressive decrease the total heat flux, as the larger inter-ion distances in the zigzag and helical configurations reduce the interaction between neighboring ions. 

The transition through a helical configuration in between two perpendicular zigzag configurations results in a local maximum of the total heat flow. Thus, the non-linear effects that arise in the dynamics during the transition to the planar zigzag configuration are more detrimental to heat transport than those corresponding to the helical configuration. 

Interesting issues to consider in future work include the study of additional transport properties, such as the conductance, and their correlation with the atomic delocalization in systems of different dimensionality and size. In the case of Coulomb crystal of trapped ions, the study of heat transport properties in systems with different ion species or with structural defects are also of great interest. 

%%%%%%%%%%%%%%%%%%%%%%%%%%%%%%%%%%%%%%%%%%%%%%%%%%%

\begin{acknowledgments}

We thank J. P. Palao and J. Onam Gonz\'alez for fruitful discussions, and J. Gonz\'alez-Platas for his guidance on the use of the VESTA software. This project was funded by the Spanish MICINN and European Union (FEDER) (FIS2017-82855-P).

\end{acknowledgments}

%%%%%%%%%%%%%%%%%%%%%%%%%%%%%%%%%%%%%%%%%%%%%%%%%%%%%%%%%%%%

\end{document}